\documentclass[twocolumn,preprintnumbers,nofootinbib,prl]{revtex4}

\usepackage{amssymb,amsmath,graphicx}
\usepackage{epsf}
\usepackage{graphicx,epsfig}
\usepackage{amsfonts}
\usepackage{amssymb}
\usepackage{xcolor}

\def\half{\frac{1}{2}}

\def\tk{{\tilde k}}

\begin{document}

\newcommand{\be}{\begin{equation}}
\newcommand{\ee}{\end{equation}}
\newcommand{\bea}{\begin{eqnarray}}
\newcommand{\eea}{\end{eqnarray}}
\newcommand{\barr}{\begin{array}}
\newcommand{\earr}{\end{array}}
\def\bal#1\eal{\begin{align}#1\end{align}}

\pagestyle{plain}


\title{Unique Fingerprints of Alternatives to Inflation in the Primordial Power Spectrum}

\author{Xingang Chen$^{1}$, Abraham Loeb$^{1}$, and Zhong-Zhi Xianyu$^{2,3}$}
\affiliation{\vspace{2mm}
$^1$Harvard-Smithsonian Center for Astrophysics, 60 Garden Street, Cambridge, MA 02138, USA\\
$^2$Center of Mathematical Sciences and Applications, Harvard University, 20 Garden Street, Cambridge, MA 02138, USA\\
$^3$Department of Physics, Harvard University, 17 Oxford Street, Cambridge, MA 02138, USA
}


\begin{abstract}
Massive fields in the primordial Universe function as standard clocks and imprint clock signals in the density perturbations that directly record the scale factor of the primordial Universe as a function of time, $a(t)$. A measurement of such signals would identify the specific scenario of the primordial Universe in a model-independent fashion. In this Letter, we introduce a new mechanism through which quantum fluctuations of massive fields function as standard clocks. The clock signals appear as scale-dependent oscillatory signals in the power spectrum of alternative scenarios to inflation.
\end{abstract}
\maketitle

{\em Introduction.}
Recent debate \cite{Ijjas:2017,SciAmReply,SciAmWeb} over cosmic inflation as the origin of the standard big bang model has sharpened the question of how to model-independently distinguish the inflation scenario \cite{Guth:1980zm} from alternative scenarios \cite{Khoury:2001wf,Gasperini:1992em,Wands:1998yp,Brandenberger:1988aj} in which the density perturbations \cite{Mukhanov:1981xt} might be produced during a non-inflationary phase.

The main criticism is that inflation may not be falsifiable as a scenario -- notice that the key point here is {\em not} whether any individual inflation model can be ruled out or not. We use the terminology ``scenario" instead of ``model" to emphasize this point. Sometimes, the process of ruling out and selecting inflation models within the framework of the inflation scenario is compared to the process of selecting the particle physics standard model within the framework of quantum field theory. However, the current status of the inflation scenario is very different from that of quantum field theory. Independent of the standard model and before the selection process, quantum field theory had already passed many experimental tests of its defining properties against alternative frameworks.

Nonetheless, inflation indeed had a falsifiable prediction as a mechanism for generating the density perturbations. Namely, the perturbations have to be seeded on superhorizon scales. This prediction was firmly verified by observations. However, it has been shown through at least toy examples \cite{Khoury:2001wf,Gasperini:1992em,Wands:1998yp,Brandenberger:1988aj} that this consequence does not have to come from an inflationary background. For example, this is possible even in certain contracting backgrounds. It is these alternative scenarios that inflation is now competing with, although they have more unsolved challenges than inflation. On the other hand, due to model-building flexibility, inflation has much more flexible predictions on values of other observables such as flatness, spectral index, adiabaticity, tensor mode and non-Gaussianity. It is not clear which specific predictions would have falsified or can be used as falsifiable predictions of the inflation scenario \cite{SciAmWeb}. This is the main criticism. The central question is how to {\em model-independently} establish any primordial scenario -- among inflation and alternative scenarios -- through experimental data.

What are the defining properties of these scenarios? And do we have potential observables that could directly reveal these properties? The answer to the first question is in fact very simple -- the evolution of the scale factor as a function of time, $a(t)$. Different scenarios have drastically different $a(t)$s. The problems arise because observables that are commonly used do not directly measure $a(t)$. The links between these observables and $a(t)$ are so indirect that different types of $a(t)$ could predict the same value for one observable. But, can we find any observable that directly measures $a(t)$? It has been shown that such observables indeed exist in principle \cite{Chen:2011zf,Chen:2014joa,Chen:2014cwa,Chen:2015lza}.

{\em Primordial standard clocks.}
Any simple model of the primordial Universe in reality needs to be embedded into a larger UV-complete framework. Despite our ignorance about the details of this framework, we do know that, whatever this theory is, it requires a tower of extra fields. Some of these fields must have masses that are larger than the characteristic energy scale of the simple model -- the scale of the event horizon. These fields oscillate like harmonic oscillators. It has been shown that these harmonic oscillations leave their imprints in the density perturbations and directly record the background $a(t)$ \cite{Chen:2011zf,Chen:2014joa,Chen:2014cwa,Chen:2015lza}.
We call these massive fields the primordial standard clocks.
The oscillations may be excited classically or quantum mechanically. In the former case \cite{Chen:2011zf,Chen:2014joa}, an extra ingredient, namely some sharp feature, is required. On the other hand, quantum excitation of standard clocks is universal, and it has been shown that the clock signals exist in all inflationary models, and possibly in many models of alternative scenarios too \cite{Chen:2015lza}. Although the amplitude is very model dependent, the pattern of the clock signal is not and directly measures $a(t)$.

Previous realizations of primordial standard clocks, including both classical and quantum, rely on the resonance mechanism between the clock field and the field seeding the density perturbations (see \cite{Chen:2016qce} for a simple review). Due to this mechanism, in the case of quantum primordial standard clocks, the clock signals hide in the shape of non-Gaussianities \cite{Chen:2015lza}.

In this Letter, we introduce a new mechanism through which quantum fluctuations of massive fields function as standard clocks.
This mechanism will be mainly applied to alternative scenarios in this work.
The clock signals manifest themselves as scale-dependent oscillatory signals in the density perturbations. It is therefore much easier to constrain or extract these from data than those shape-dependent ones in non-Gaussianities.

{\em Clock signals in massive fields.}
We use the following simple parametrization to characterize the background of any arbitrary primordial Universe scenario,
\bea
a(t) = a_0 \left( \frac{t}{t_0} \right)^p
= a_0 \left( \frac{\tau}{\tau_0} \right)^{\frac{p}{1-p}} ~,
~
\tau = \frac{1}{1-p} \frac{t_0^p}{a_0} t^{1-p} ~,
\label{a(t)ansatz}
\eea
where $t$ and $\tau$ are physical and conformal time in the FRW metric, respectively.
The type of scenario is uniquely determined by the value of $p$ \cite{Chen:2011zf}. Inflation scenario corresponds to $|p|>1$ (namely, $|\epsilon|<1$). Contraction scenarios correspond to $0<p<1$. Non-inflationary expansion scenarios correspond to $-1<p<0$. $t$ runs from 0 to $+\infty$ (with $t_0>0$) for $p>1$, and from $-\infty$ to 0 (with $t_0<0$) for all other $p$ values. $\tau$ always runs from $-\infty$ to 0 (with $\tau_0<0$). Such a background has an event horizon with energy scale $m_{\rm h} \equiv |(1-p)/t|$.

In the background (\ref{a(t)ansatz}), the equation of motion (EOM) for the mode function (denoted as $v_k$) of a field with mass $m$ and wave number $k$ is
\bea
\ddot v_k + \frac{3p}{t} \dot v_k + \frac{k^2}{a^2} v_k + m^2 v_k =0 ~,
\label{eom_vk}
\eea
where a dot denotes a derivative with respect to $t$.

The mode function $v_k$ may evolve through three different regimes, determined by which of the last three terms in (\ref{eom_vk}) dominates. The regime dominated by the mass term is called the classical regime \cite{Chen:2015lza}. When the momentum term dominates, the mode enters the $k$-dominated regime; and when the second term of (\ref{eom_vk}) dominates, the mode is in the horizon-dominated regime.

\begin{figure}[t]
  \centering
  \includegraphics[width=0.52\textwidth]{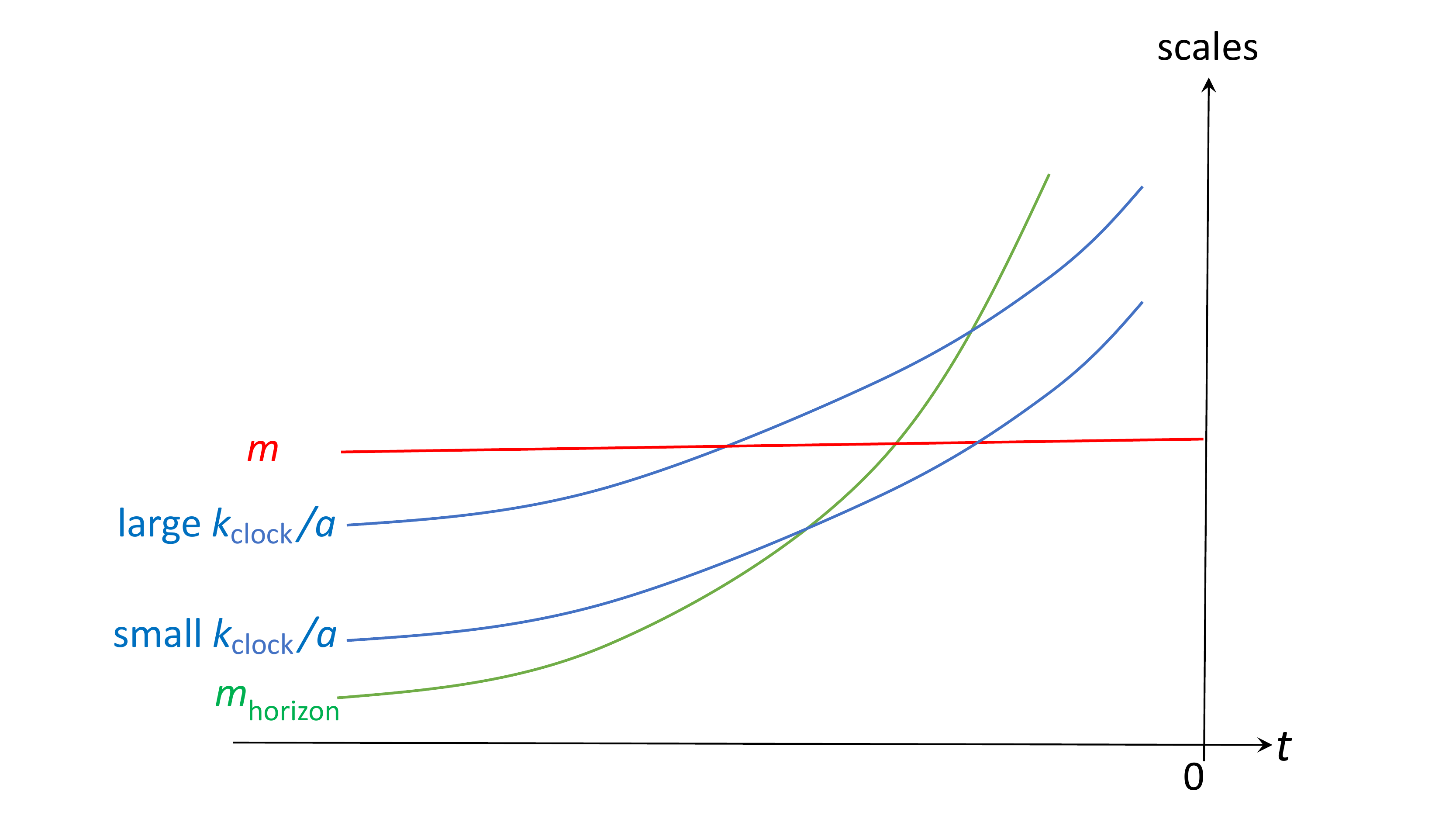}
\caption{\small Sketch of the three scales in contraction scenarios, including the difference between the small and large $k$ cases.}
\label{Fig:small_large_k}
\end{figure}

We now use contraction scenarios ($0<p<1$), such as Ekpyrotic \cite{Khoury:2001wf} and matter bounce \cite{Wands:1998yp}, as examples. The three regimes are illustrated in Fig.\;\ref{Fig:small_large_k}.

At early time the mass term dominates, and the EOM behaves approximately as
\bea
\ddot v_k + m^2 v_k \approx 0 ~.
\eea

If $k$ is sufficiently large to satisfy
\bea
\tk
\equiv \frac{k/a}{m_{\rm h}}
\left( \frac{m_{\rm h}}{m} \right)^{1-p}
> 1 ~,
\label{large_k_cond}
\eea
then there exists a period of time during which the $k$-term dominates and the EOM reads approximately,
\bea
v''_k + k^2 v_k \approx 0 ~,
\eea
where a prime denotes a derivative with respect to $\tau$.
For smaller $k$, the $k$-term is never important. Note that the condition (\ref{large_k_cond}) is time independent.

At late time, the 2nd term in Eq.~(\ref{eom_vk}) will always become more important than the last two terms. Physically this means that the horizon shrinks faster than the scale factor and perturbations eventually exit the horizon despite the contracting background.

We now show an important property of the massive field mode function when condition (\ref{large_k_cond}) is satisfied.
During the transition from the classical regime to the $k$-dominated regime, the mode function develops a $k$-dependent phase which directly records the evolution of the scale factor $a(t)$.

In these two regimes, the EOM is approximately,
\bea
\ddot v_k + \frac{k^2}{a^2} v_k + m^2 v_k \approx 0 ~.
\label{eom_vk_classical_k-dominated}
\eea
Neglecting an envelope with weaker time dependence, the leading order solution to this equation is
\bea
v_k \sim \exp \left[ \pm im \int_{y_0}^y \sqrt{\frac{1}{z^2}+1}
\frac{ d a^{(-1)}\left( \frac{kz}{m} \right) }{dz} dz
\right] ~,
\label{vk_approx}
\eea
where we have defined $y\equiv ma(t)/k$ and $y_0$ is evaluated at $t_0$ in the far past. $a^{(-1)}(kz/m)$ denotes the inverse function of $a(t)$. We can already see how the inverse function of $a(t)$ gets encoded in the phase as a function of $k$.
This does not depend on how we parameterize $a(t)$.
More explicitly, we apply the parametrization (\ref{a(t)ansatz}) and get
\bea
v_k \sim \exp \left[
\pm i \frac{t_0 m}{p a_0^{1/p}} \left( \frac{k}{m} \right)^{1/p}
\int_{y_0}^y \sqrt{\frac{1}{z^2}+1} z^{\frac{1}{p}-1} dz
\right] ~.
\label{vk_in_y}
\eea
In the far past (large $y$), we find $v_k \propto \exp\left( \pm imt \right)$. This is the expected mass-dominated behavior in a contraction scenario. In the $k$-dominated regime (small $y$), using the same normalization for $v_k$, we find
\bal
v_k &\propto
\exp \left( \pm i \beta \tk^{1/p} \right)
\exp \left(\pm ik\tau \right)
~,
\label{vk_clock_signal}
\eal
where $ \beta = -\Gamma\left( \frac{1}{2p} + \half\right)
\Gamma\left( -\frac{1}{2p} \right)/(2\sqrt{\pi})$.

The $k$-dependence in the first phase factor in (\ref{vk_clock_signal}) is given by the inverse function of $a(t)\propto t^p$. Recall that the value of $p$ determines the scenario. We will make use of this important property when computing the density perturbations. As in previous works, we label this phase factor and related signal as the ``clock signal".

Before proceeding to the analyses of the density perturbations, a few comments about the generality and robustness of the clock signals are in order:

\noindent (i) Caution should be exercised in deriving the precise values of $\beta$. $\beta$ becomes singular when $p=1/(2n)$ with $n$ a positive integer. The singularity is removed by higher order terms in the series expansion of the resulting integral in (\ref{vk_in_y}) around $y\to \infty$. For example, with $p=1/2$, after the cancellation we find that $\beta$ should be replaced by $1/4 + (\ln2)/2 - (\ln\tk)/2$, where the $\ln\tk$ term is due to normalizing a phase-shift at large $y$. Higher order terms are also important if $p$ is close to those singularities.

\noindent (ii) The results above for contraction scenarios can be generalized to other scenarios with slight modifications.

For expansion alternative scenarios ($-1<p<0$), the massive field starts from the $k$-dominated regime and then enters the classical regime. The condition (\ref{large_k_cond}) should then be changed to
\bea
\tilde k <1 ~.
\label{small_k_cond}
\eea
Also, $y_0\ll 1$ and $y_0 <y$, instead of $y_0\gg 1$ and $y_0 >y$.
Consequently, $v_k$ initially behaves as
$ v_k \propto \exp(\pm ik\tau)$, and develops a $k$-dependent phase after entering the classical regime,
\bea
v_k \propto
\exp \left( \mp i \beta \tk^{1/p} \right)
\exp \left(\pm imt \right)
~.
\label{vk_expansion_scenarios}
\eea

For inflation models with $|p|\gg 1$, $m_{\rm h}$ never exceeds $m$, so the clock signal phase factor is present for all $k$. In the limit of the exponential inflation $|p|\to \infty$, (\ref{vk_expansion_scenarios}) becomes
\bea
v_k \propto \exp\left[
\mp \frac{im}{H}
\ln \left(-\frac{H}{m} k\tau \right)
\right] ~,
\label{vk_clock_signal_inflation}
\eea
where we have used the relation (\ref{a(t)ansatz})
and $p/t_0 \to H$. The dependence of the phase on $k$ is still the inverse function of $a(t)$; however, it depends on the combination $k\tau$ which has special implications for this case.

\noindent (iii) Although mathematically the mode function can contain an arbitrary $k$-dependent phase factor, physically such factors are constrained by initial conditions. For contraction scenarios, the ground state of the initial vacuum fluctuations of massive fields is $k$-independent harmonic oscillations; for expansion scenarios, it is effectively massless plane waves. One can also consider various physically motivated excited states, but it is clear that a special initial state just able to cancel the very special form of the clock signal in (\ref{vk_clock_signal}) or (\ref{vk_expansion_scenarios}), is highly artificial.

\noindent (iv) We examine whether the clock signal may be contaminated during the subsequent evolution of $v_k$.

For contraction scenarios, the horizon scale $m_{\rm h}$ will eventually dominate. The transition from the $k$-dominated to the horizon-dominated regime is described approximately by (\ref{eom_vk}) with the last term neglected.
This is the same as the EOM of the massless mode and the solutions are Hankel functions of $k\tau$ for arbitrary $p$.
The $k$-dependent factor generated in the transition can be seen by taking the limit $k\tau \to 0$. It is in power-law in $k$ and not important for our purpose.

For expansion scenarios, the massive mode either stays in the classical regime (for the inflation scenario), so the clock signal remains the same; or transitions from the classical to the horizon-dominated regime (for alternative scenarios), in which case the $k$-term is negligible, so there is no extra $k$-dependent term generated.

Therefore, in all cases the clock signal generated in the transition from the classical to the $k$-dominated regime (or the reverse) will not change during the subsequent evolution of the massive field.

The clock signal is always generated in the subhorizon quantum regime. The quantum-to-classical transition for each mode after horizon exit, carrying the clock signal phase factor, works in the same way as shown in Ref.~\cite{Polarski:1995jg}.

\noindent (v) So far, approximations are used to extract the clock signal because the EOM (\ref{eom_vk}) and (\ref{eom_vk_classical_k-dominated}) cannot be solved analytically for arbitrary $p$. Nonetheless, a few special cases with analytical solutions are available for us to check our results:

a) $v_k$ with $p=1/2$ is analytically solvable. With the initial condition $v_k\propto e^{-imt}$, $v_k$ has a phase factor
$\propto \exp \left[ i \left(-1/4-(\ln 2)/2+ (\ln\tk)/2 \right) \tk^2 \right]$ at late time. This matches what we found above.

b) $p=1/3$ is solvable analytically when neglecting the horizon term. This approximation is fine because, as discussed, the horizon term does not change the clock signal. Imposing
$v_k\propto e^{-imt}$ as initial condition, in the $k$-dominated regime we find a phase factor $\propto \exp(\frac{2i}{3}\tk^3)$, agreeing with (\ref{vk_clock_signal}).

c) The exponential inflation case is also analytically solvable. At late time the massive field has a phase factor $\propto (k\tau)^{-i m/H}$ if we impose the Bunch-Davies initial condition and consider $m\gg H$. This agrees with (\ref{vk_clock_signal_inflation}).

For general cases, one can always perform numerical calculations to find the forms of clock signals.

\begin{figure}[h]
  \centering
  \includegraphics[width=0.40\textwidth]{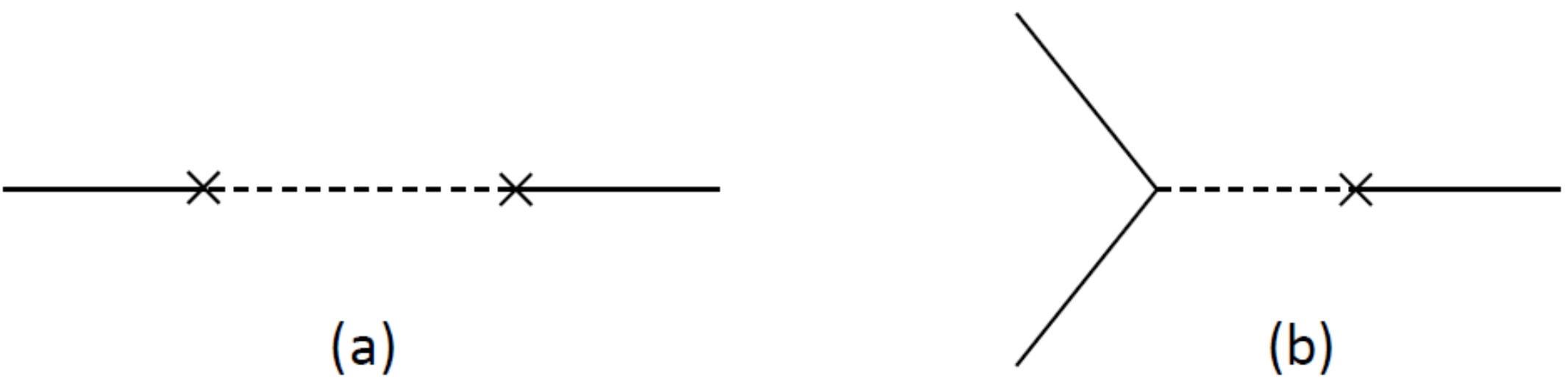}
\caption{\small Examples of the massive field contribution to the (a) two-point and (b) three-point correlation functions. The solid line represents the massless curvature perturbation $\zeta$, and the dashed line represents the massive clock field. Both diagrams belong to the so-called single-exchange diagrams.}
\label{Fig:2pt_3pt_vertices}
\end{figure}

{\em Clock signals in density perturbations.}
Now we consider the correction from massive field to the two-point correlation of curvature perturbation $\zeta$ (with mode function $u_k$) in Fig.~\ref{Fig:2pt_3pt_vertices}. We take a direct coupling between the two modes for simplicity with a general time-dependent but non-oscillatory coupling strength $\lambda_2(\tau)$. One can choose other types of couplings without changing the main results here. Then Fig.~\ref{Fig:2pt_3pt_vertices}(a) reads
\bal
\label{ininint}
&\Delta \langle \zeta^2 \rangle'
=
2 u_k^* u_k|_{\tau_{\rm end}}
\left|
\int_{-\infty}^{\tau_{\rm end}} d\tau \lambda_2(\tau) u_k v_k
\right|^2
\nonumber \\
& -4 {\rm Re} \left[
u_k^2|_{\tau_{\rm end}}
\int_{-\infty}^{\tau_{\rm end}}\!\!\! d\tau_1 \lambda_2(\tau_1) u^*_k v_k
\int_{-\infty}^{\tau_1} \!\!\! d\tau_2\lambda_2(\tau_2) u^*_k v_k^*
\right].
\eal
The prime on $\Delta \langle \zeta^2 \rangle$ means that the momentum conservation delta function is removed.

In the classical and $k$-dominated regimes, $v_k$ rapidly oscillates in time, and so the integrand does not contribute much to the integral. Resonance between different oscillatory components may give larger contributions, but this requires at least two modes with different wave numbers (for reasons, see \cite{Chen:2016qce}). Consequently, in previous realizations of primordial standard clocks, either some background oscillation \cite{Chen:2011zf} or soft limits of non-Gaussian correlations \cite{Chen:2015lza} were required.

However, to study the scale dependence of the correlation functions, all mode functions share the same wave number and the resonance does not occur. For alternative scenarios, the main contribution to the integrals in (\ref{ininint}) originates from the horizon-dominated epoch, during which both $v_k$ and $u_k$ stop oscillating. As we have shown, $v_k$ has developed a phase factor -- the clock signal -- until this epoch. This phase is time-independent and can be pulled out of the integral. It is then straightforward to figure out the clock signals in various correlation functions.

\begin{figure}[t]
  \centering
  \includegraphics[width=0.45\textwidth]{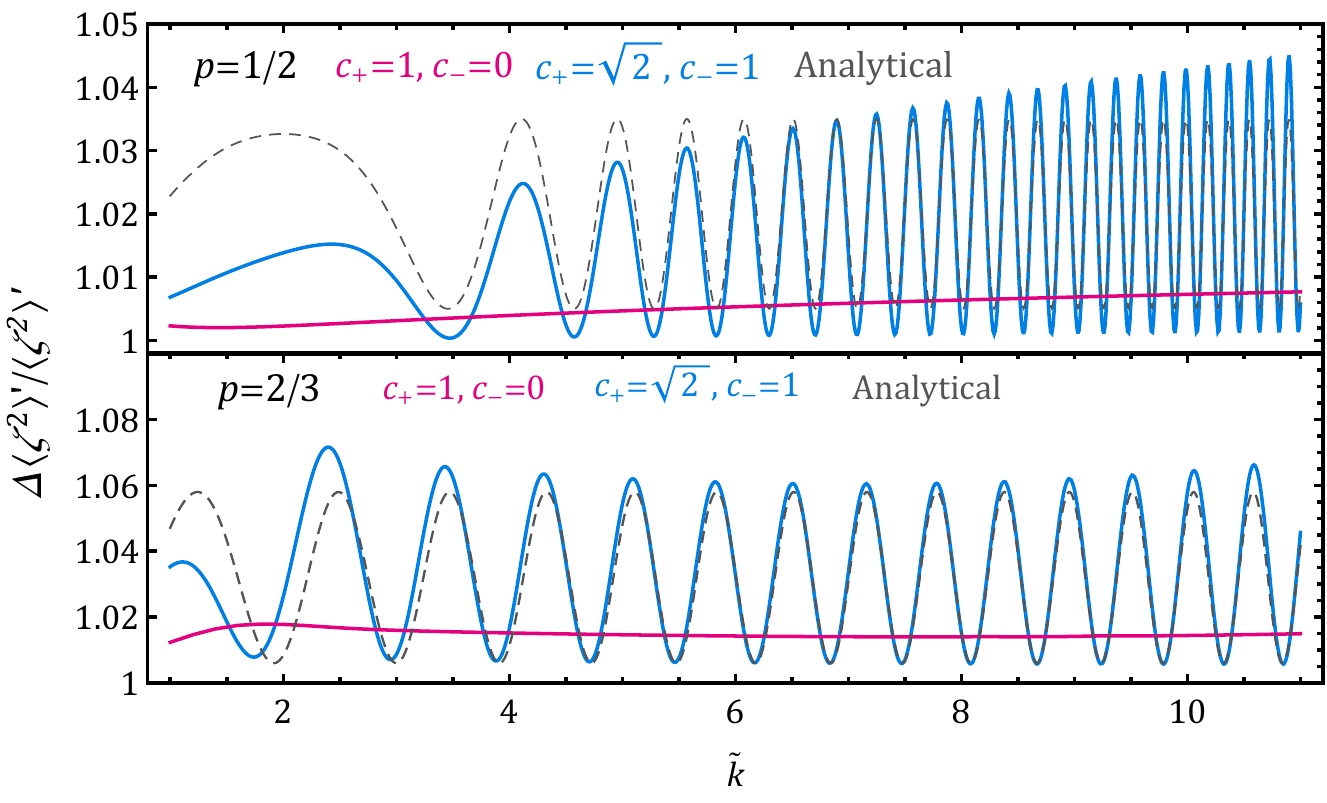}
\caption{\small Corrections to scalar power spectrum from a massive field for two contraction scenarios. In each panel, two numerical results (described by blue and magenta solid lines) with different initial conditions are plotted against the analytical clock signals in (\ref{2pt_clock_signal}) (dashed). In numerical calculations, the EOM approach \cite{Chen:2015dga} is applied and we use the examples $\lambda_2 = 0.05 a(\tau)$ ($p=1/2$) and $\lambda_2 = 0.02 a(\tau)$ ($p=2/3$) for faster convergence. The mismatch in low $\tilde k$ is because the condition $\tilde k \gg 1$ is needed for Eq.~(\ref{eom_vk_classical_k-dominated}) to be a good approximation.}
\label{Fig:clock_numerical}
\end{figure}

For contraction scenarios, if we start the massive field as $v_k \sim c_+ e^{-imt} + c_- e^{imt}$, then until the horizon-dominated regime these two components will evolve into the two in (\ref{vk_clock_signal}). This is the regime where the integral in (\ref{ininint})  receives the main contribution. So the two-point correlation function contains a $k$-dependent oscillatory component, namely the clock signal, as follows,
\bal
\Delta \langle \zeta^2 \rangle'
=
|c_+ c_-^*| \sin \left( 2\beta\tk^{1/p} + \phi \right)
f(k)
+ \dots
~,
\label{2pt_clock_signal}
\eal
where both $c_\pm$ and $\phi$ are approximately constants. We use $f(k)$ to denote the envelope and dots to denote other terms, which have much weaker $k$-dependence than the oscillations in the clock signal. The pattern of the clock signal (parameterized by $p$) is the unique prediction of a particular scenario.
Measuring such a pattern would directly determine $a(t)$.
Two examples are plotted in Fig.~\ref{Fig:clock_numerical} in comparison with numerical results \cite{footnote_figs}.
On the other hand, the amplitude of the clock signal depends on couplings and does not belong to model-independent predictions of any particular scenario.

Expansion scenarios are similar except for the case of exponential inflation ($|p|\to \infty$), for which due to the special form of the mode function (\ref{vk_clock_signal_inflation}), the $k$-dependent clock signal gets cancelled after integration over time.
This is possible because, unlike the alternative scenarios, the massive field never exits the classical regime and keeps oscillating rapidly.
The cancellation can be easily seen by redefining the integration variable as $k\tau$. This result is also expected from the scale invariance of the exponential inflation. With large but finite $p$, the scale-dependent clock signal is present but suppressed by $1/p$. For nearly exponential inflation, we should look for quantum standard clock signals in the shape of non-Gaussianities instead \cite{Chen:2015lza}.

Vacuum states are in general more complicated in alternative scenarios than in inflation. In some scenarios the thermal state is required to achieve a scale-invariant spectrum \cite{Brandenberger:1988aj}. For generality, we have kept both positive and negative frequency components in the massive field initial state. Both are required for the leading order clock signal to show up as scale-dependent corrections in density perturbations.

Next we consider the three-point correlation in Fig.~\ref{Fig:2pt_3pt_vertices}(b). The appearance of the intermediate massive field is the same as before, except that now we have permutation of the three momenta. We leave the general configuration for future studies. For the purpose of this work, the overall scale dependence of non-Gaussianity can be studied by looking at the equilateral configuration where $k_1=k_2=k_3\equiv k$. The scale dependence as a function of $k$ is exactly the same as in (\ref{2pt_clock_signal}). Notice that this clock signal is not only correlated with the signal in the power spectrum (\ref{2pt_clock_signal}), but also with the shape-dependent clock signal in the same three-point correlation studied in \cite{Chen:2015lza}. These correlations distinguish the standard clocks from other models (such as models with periodic features) which may mimic the signals in the power spectrum.

{\em Experimental prospects.}
So far, there is no evidence for primordial features in the cosmic microwave background (CMB). The best constraints come from the Planck data \cite{Akrami:2018odb} and sensitively depend on the location and frequency of the features in the $\ell$-space. Nonetheless, there are a couple of interesting candidates. In $\ell\sim 20$ - $30$, there is a well-known sharp feature candidate \cite{Peiris:2003ff} with best-fit amplitude $\Delta P_\zeta/P_\zeta \sim 0.2$. Around $\ell \sim 700$ - $800$, there is another feature candidate \cite{Akrami:2018odb} that could have several possible origins \cite{Chen:2014cwa}. The possibilities include the sharp feature model, resonance model, or primordial standard clock model. The last category includes signals studied in this work. The best-fit amplitude is around $\Delta P_\zeta/P_\zeta \sim 0.03$ - $0.05$.

Currently both candidates are consistent with statistical fluctuations. The constraints are expected to be improved by polarization data of future CMB experiments, and more significantly, by large-scale structure (LSS) surveys due to their 3D information \cite{Huang:2012mr,Chen:2016vvw,Ballardini:2016hpi}.
Very futuristically, the 21 cm tomography may be used to search for feature models in much shorter scales with high precisions \cite{Chen:2016zuu,Xu:2016kwz}.

Here we forecast the constraints on the quantum primordial standard clock signals of this work from the near future photometric LSS survey, such as the LSST \cite{Abell:2009aa}. The methodology and the specification of the experiment follow Ref.~\cite{Chen:2016vvw}. We use the following template to capture the main properties of our results,
\bea
\frac{\Delta P_\zeta}{P_\zeta} =
C \sin \left[ p \Omega \left( \frac{2k}{k_r} \right)^{1/p} + \phi \right] ~,
\label{clock_template}
\eea
where $C$, $\Omega$, $k_r$ and $\phi$ are all constants. The fiducial values are chosen to approximately represent those of the best-fit candidate around $\ell\sim 700$ - $800$, as well as values around them. The results are in Table.~\ref{Table:LSSforcast}. We can see that the errors on the amplitude will be improved by one order of magnitude, and, in case of detection, the value of $p$ can be determined very precisely due to the oscillatory nature of the clock signal.

\begin{table}
  \begin{center}
\begin{tabular}{|c||c|c|c|c|c|c|}
\cline{2-7}
\multicolumn{1}{c|}{}& {\footnotesize $p=2/3$} & {\footnotesize $p=2/3$} & {\footnotesize $p=2/3$} & {\footnotesize $p=1/5$} & {\footnotesize $p=1/5$} & {\footnotesize $p=1/5$} \\
\multicolumn{1}{c|}{}& {\footnotesize $\Omega=30$} & {\footnotesize $\Omega=30$} & {\footnotesize $\Omega=100$} & {\footnotesize $\Omega=30$} & {\footnotesize $\Omega=30$} & {\footnotesize $\Omega=60$} \\
\multicolumn{1}{c|}{}& {\footnotesize $k_r=0.1$} & {\footnotesize $k_r=0.2$} & {\footnotesize $k_r=0.1$} & {\footnotesize $k_r=0.1$} & {\footnotesize $k_r=0.2$} & {\footnotesize $k_r=0.1$} \\
\hline
 {\footnotesize $\sigma_C$} & {\footnotesize 0.0016} & {\footnotesize 0.0016} & {\footnotesize 0.0016} & {\footnotesize 0.0016} & {\footnotesize 0.0017} & {\footnotesize 0.0016} \\ \hline
 {\footnotesize $\sigma_p$} & {\footnotesize 0.0012} & {\footnotesize 0.0033} & {\footnotesize 0.00035} & {\footnotesize $1.3\times 10^{-6}$} & {\footnotesize $3.8\times 10^{-5}$} & {\footnotesize $5.6\times 10^{-7}$} \\ \hline
\end{tabular}
\caption{\small Marginalized $1$-$\sigma$ constraints on the parameters of the clock signal template \eqref{clock_template} of two contraction scenarios from a LSST-like survey. The other fiducial values not presented in the table are $C=0.03$ and $\phi=0$. $k_r$ is in unit of Mpc$^{-1}$. Note that $k_r$ and $\Omega$ are degenerate in the template. The values $C=0.03$, $\Omega=30$, $k_r=0.1$ approximately represent the best-fit values of the candidate around $\ell\sim 700$ - $800$.}
\label{Table:LSSforcast}
  \end{center}
\end{table}

{\em Discussion.}
Could the clock signals (\ref{2pt_clock_signal}) in alternative scenarios be mimicked by inflation models?
Scale-dependent oscillatory features may be generated during inflation if there are disturbances to the attractor solution, which may be due to new physics at short distances (see \cite{Brandenberger:2012aj} for a review), or sharp or periodic features in the model (see \cite{Chen:2010xka} for reviews). Depending on whether the disturbances are generated for all modes at the same time or at the same energy scale, the power spectrum acquires an oscillatory component that behaves as $\sim \sin(k/k_0 + \phi)$ or $\sim \sin(\Omega \log(k) + \phi)$, where $k_0$, $\Omega$ and $\phi$ are all constants. These are distinctively different from (\ref{2pt_clock_signal}).

To mimic the clock signal (\ref{2pt_clock_signal}), one may construct a series of non-periodic features on the inflation potential with just the right spacings to reproduce the oscillatory patterns in (\ref{2pt_clock_signal}). Comparing to the simple physics of one massive state in an alternative scenario, this finely tuned procedure is highly artificial with many extra parameters. It would be clearly disfavored even without a complete understanding of the subtle measure problem in inflation. The differences become more clear if non-Gaussianities can be measured.
As mentioned, the clock signal (\ref{2pt_clock_signal}) is correlated with corresponding clock signals residing in both the scale and shape dependence of non-Gaussianities. On the other hand, the inflation model with artificial features also has correlated features in non-Gaussianities \cite{Chen:2010xka}. It is easy to see that the two are very different.

How density perturbations in contraction scenarios survive the bounce remains an open question \cite{Brandenberger:2016vhg}. However, since the most important pattern of the clock signal is unchanged under a universal rescaling of all wave numbers or the amplitude, the information about $a(t)$ in the clock signal most likely remains unchanged under whatever bounce mechanism is at work.

\begin{figure}[t]
  \centering
  \includegraphics[width=0.42\textwidth]{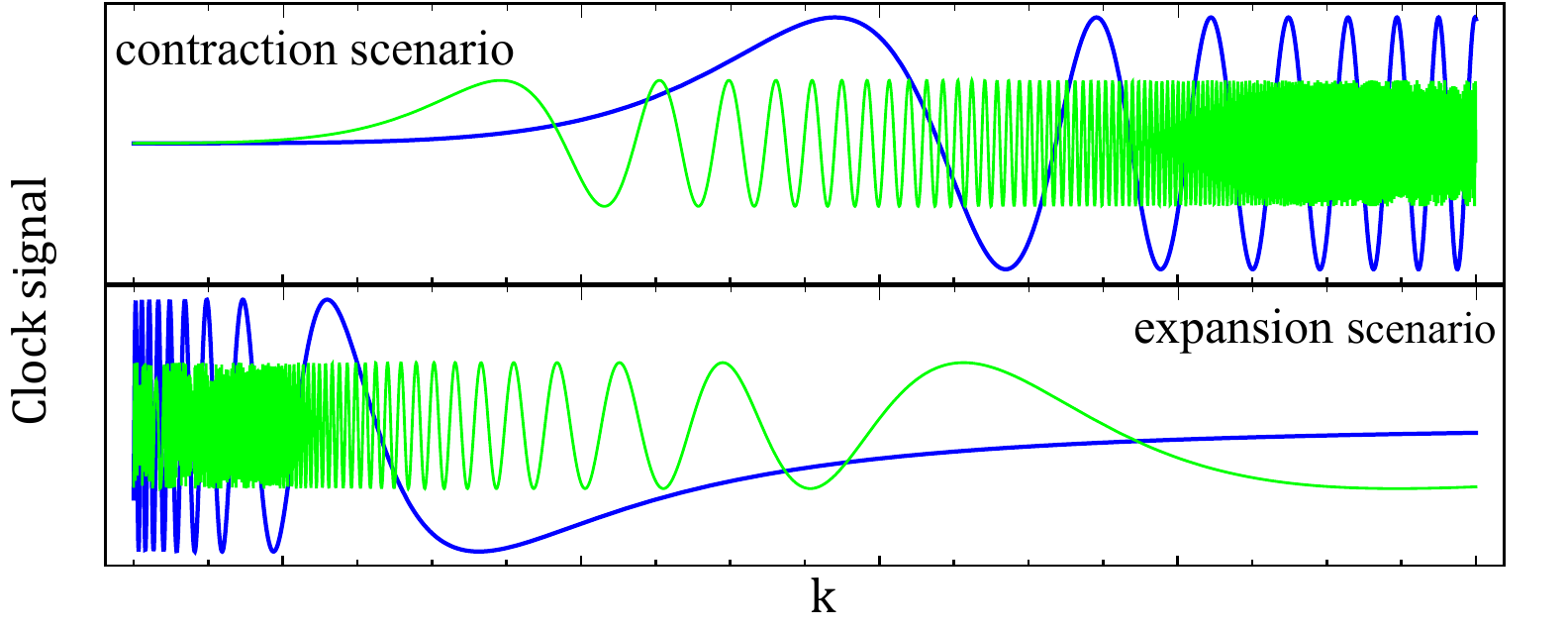}
\caption{\small Illustration of clock signals for two types of alternative scenarios - expansion and contraction scenarios. Two different masses are plotted in each case.}
\label{Fig:clock_signals}
\end{figure}

We conclude by making the following interesting comparison between the inflation and alternative scenarios.
The energy scale of horizon during inflation is roughly a constant. Particles with a mass around or below this scale are created quantum mechanically and leave their imprints in the shapes of non-Gaussianities of the density perturbations. In this sense, inflation works like a particle collider with a fixed energy scale \cite{Chen:2009we,Baumann:2011nk,Noumi:2012vr,Gong:2013sma,Kehagias:2015jha, Arkani-Hamed:2015bza,Dimastrogiovanni:2015pla,Lee:2016vti,Meerburg:2016zdz, Chen:2016uwp,Kehagias:2017cym,An:2017hlx,Iyer:2017qzw, Kumar:2017ecc,MoradinezhadDizgah:2018ssw,Tong:2018tqf,Saito:2018omt}.

In contrast, the energy scale of the horizon in alternative scenarios increases with time. Fields with a mass much larger than the horizon scale (and hence more difficult to create spontaneously) may eventually become lighter (hence easier to create). These are the fields of interest here. We have shown that with some generic assumptions on the initial state, they leave imprints in the density perturbations in terms of scale-dependent oscillatory signals. These signals appear in wave numbers that satisfy condition (\ref{large_k_cond}) or (\ref{small_k_cond}) for contraction or expansion scenarios, respectively (see Fig.~\ref{Fig:clock_signals}). Therefore, the alternative scenarios are more like particle scanners -- they scan over a tower of massive fields one by one and display each of them as a pulse of signals at different length scales in the density perturbations.

Besides particle spectra, clock signals from both types of particle detectors also carry direct information about $a(t)$, and therefore are predictions which can be used to falsify competing primordial Universe scenarios in a model-independent fashion.

{\em Acknowledgment.}
We would like to thank Hayden Lee, Chon-Man Sou and Matias Zaldarriaga for helpful discussions, and Mohammad Hossein Namjoo for assistance in LSS forecast. AL was supported in part by the Black Hole Initiative at Harvard University, which is funded by a JTF grant. ZZX is supported in part by Center of Mathematical Sciences and Applications, Harvard University.


\begin{thebibliography}{99}
\bibitem{footnote_figs}
The single-field examples of alternative scenarios do not give a scale-invariant $\langle\zeta^2\rangle$ except for $p=2/3$; for other values of $p$, additional fields are required. These model-dependent aspects do not affect the clock signal factor in (\ref{2pt_clock_signal}). Therefore, in Fig.~\ref{Fig:clock_numerical}, we plot $\Delta\langle\zeta^2\rangle'/\langle\zeta^2\rangle'$. In Fig.~\ref{Fig:clock_signals}, we only plot the clock signal.
\end{thebibliography}
\end{document}